\documentclass[12pt]{spieman}  
\usepackage{amsmath,amsfonts,amssymb}
\usepackage{graphicx}
\usepackage{setspace}
\usepackage{tocloft}

\title{Schedule optimization for transiting exoplanet observations with NASA's Pandora SmallSat mission}

\author[a,*]{Trevor O. Foote}
\author[b,c]{Thomas Barclay}
\author[b,c]{Christina L. Hedges}
\author[a]{Nikole K. Lewis}
\author[b]{Elisa V. Quintana}
\author[d]{Benjamin V. Rackham}

\affil[a]{Cornell University, Department of Astronomy and Carl Sagan Institute, Ithaca, New York, United States}
\affil[b]{NASA Goddard Space Flight Center, Greenbelt, Maryland, United States}
\affil[c]{University of Maryland, Baltimore County, 1000 Hilltop Circle, Baltimore, Maryland, United States}
\affil[d]{Massachusetts Institute of Technology, 77 Massachusetts Ave, Cambridge, Massachusetts, United States}

\cftpagenumbersoff{figure}
\cftpagenumbersoff{table} 
\begin{document} 
\maketitle

\begin{abstract}
Pandora is an upcoming NASA SmallSat mission that will observe transiting exoplanets to study their atmospheres and the variability of their host stars. Efficient mission planning is critical for maximizing the science achieved with the year-long primary mission. To this end, we have developed a scheduler based on a metaheuristic algorithm that is focused on tackling the unique challenges of time-constrained observing missions, like Pandora. Our scheduling algorithm combines a minimum transit requirement metric, which ensures we meet observational requirements, with a `quality' metric that considers three factors to determine the scientific quality of each observation window around an exoplanet transit (defined as a visit). These three factors are: observing efficiency during a visit, the amount of the transit captured by the telescope during a visit, and how much of the transit captured is contaminated by a coincidental passing of the observatory through the South Atlantic Anomaly. The importance of each of these factors can be adjusted based on the needs or preferences of the science team. Utilizing this schedule optimizer, we develop and compare a few schedules with differing factor weights for the Pandora SmallSat mission, illustrating trade-offs that should be considered between the three quality factors. We also find that under all scenarios probed, Pandora will not only be able to achieve its observational requirements using the planets on the notional target list but will do so with significant time remaining  for ancillary science.

\end{abstract}


{\noindent \footnotesize\textbf{*}Trevor Foote, \linkable{tof2@cornell.edu} }

\begin{spacing}{2}   

\section{Introduction} \label{sec:intro}
Scheduling is a critical component of astronomical observatory operations. It involves planning and coordination to ensure that the observing time is used effectively and that scientific goals are achieved. A typical observatory schedule can be broken down into two major categories of activities: science and operations. The science activities generally include observations of various celestial objects, such as stars, planets, galaxies, or other phenomena that are important to the scientific goals of the observatory. The operations activities tackle the logistical tasks required to achieve the science activities. Specifically for space-based observatories, operations activities include a broad range of tasks like station keeping, momentum management, ground station communication, and slewing maneuvers to move between science targets. 

As varied as the mission goals are for different observatories, so too are their schedule-building approaches. Large space-based observatories like {\it HST} and {\it JWST} are built to be multi-purpose facilities with a suite of instruments used to cover a broad range of scientific goals. This leads to schedule designs primarily focusing on immutable factors like observing efficiency \cite{Johnston_1993, Kinzel_2010}. The more science-case-specific requirements, like determining the best transit to observe during a year, are left for the individual observer to develop and provide manually. However, for small satellite (SmallSat) missions, the scientific goals are often highly focused on just one aspect of astronomical inquiry. This allows the schedule to similarly be specialized in order to maximize the scientific return based on unique metrics of the scientific study.

One of the most challenging types of astronomical observations to schedule are transiting exoplanets. Due to their time-constrained nature, capturing planetary transits, i.e. when a planet moves in front of its host star, and planetary eclipses, i.e. when the planet moves behind its host star, become `fence posts' within the schedule, which then need to be filled in around with less time-constrained observations. For a mission dedicated to observing transiting exoplanets, this makes the schedule development all the more challenging. Rather than just having some fence posts in the schedule that other observations have to be scheduled around, all scheduled observations are time-critical fence posts. 

This time-constraint challenge is highlighted by Morales et al. \cite{Morales_2022} in developing a scheduler for ESA's upcoming transit survey mission \textit{Ariel}. For the transiting-exoplanet-focused NASA SmallSat mission Pandora, the scheduling challenge is exacerbated further due to operational position differences. \textit{Ariel} is planned to be located at the Sun--Earth Lagrangian point L2, whereas Pandora will be in a Sun-synchronous, low-Earth polar orbit (LEO), located around 550~km above Earth ($\sim95.5$~minute orbital period). Being in LEO adds complexity with additional orbital constraints that come from the rapidly changing field of view of the telescope.

In this paper, we present the factors and processes used for producing a year-long observing schedule for the Pandora SmallSat. The current scheduler algorithm is a preliminary design built with minimal consideration for spacecraft operational activities and their timing. Instead, the primary focus herein is on how to optimize the scientific yield of each scheduled observation. We begin with Sec.~\ref{sec:background}, providing a brief background on exoplanet transmission spectroscopy and the stellar contamination challenge this technique currently faces, which Pandora aims to help address. Next we provide an overview of the Pandora mission in Sec.~\ref{sec:Pandora}, including the development process for the notional science target list. The methodology for developing the schedule is then broken down into three subsections: first describing how we determine the target's visibility (Sec.~\ref{subsec:target_vis}), then how we quantify the science data quality with a quality metric (Sec.~\ref{subsec:qfactor}), and finally how we ensure the schedule includes the minimum number of transits for each target, which is necessary to fulfill the mission's observational requirements (Sec.~\ref{subsec:MTRM}). Finally, in Sec.~\ref{sec:results} we present three different schedules, which were developed with our scheduling algorithm using different optimizing strategies, and discuss the trade-offs between the schedules.

\subsection{Transmission Spectroscopy} \label{sec:background}
One of the primary methods used to study exoplanet atmospheres is transmission spectroscopy \cite{Seager_2000, Brown_2001}. The use of this technique requires the exoplanet to pass in front of its host star from the viewpoint of the observer; this is called a transit. During a transit, the planet itself blocks a portion of the host star's light from reaching the observer, causing a dip in the overall brightness of the system. Planets with atmospheres block additional light as different opacity sources in the atmosphere absorb and deflect stellar light that passes through the planet's atmosphere. However, unlike light blocked by the planet, the light blocked by the atmosphere is strongly wavelength dependent. Thus, through analysis of the spectrum of the stellar light both before and during transit, we are able to infer the chemical composition and physical properties of the exoplanet's atmosphere \cite[e.g.]{Seager_2000, Lewis_2014, Kreidberg_2015}. 

This technique has already proven to be an incredibly powerful tool with {\it HST} as well as ground-based observatories \cite[e.g.]{Charbonneau_2002, Redfield_2008, Fraine_2014, Sing_2016}. Unfortunately, due to limitations in their achievable precision, these observatories have largely been constrained to characterizing only the largest planets in the exoplanet catalog, like gas giants and Neptune-sized planets. 
The improved precision of {\it JWST}, however, allows us access to smaller planets, including some Earth-sized exoplanets. Even with {\it JWST} though, smaller planets present a challenge in achieving high enough signal-to-noise to characterize their atmospheres due to the relatively thin atmospheres they possess compared to larger gas giants. One way to address this challenge is to take advantage of what has been referred to as the `small star opportunity'---that is, focus on exoplanets that orbit smaller stars, thus increasing the planet-to-star size ratio and boosting the signal-to-noise for a given planet \cite[e.g.]{Nutzman_2008, Charbonneau_2009, Dressing_2015, Kaltenegger_2020, Triaud_2021}. That means our best opportunities for characterizing sub-Neptune exoplanets in the era of {\it JWST} will be those found around smaller, cooler stars like M-dwarfs \cite[e.g.]{Seager_2009, Barstow.Irwin_2016, Morley_2017, Saidel_2020}.

One problem that arises with smaller stars is that they are more active. This leads to a higher presence of starspots, i.e. regions on the surface of a star with suppressed convection due to a stronger magnetic field, resulting in a lower temperature and reduced luminosity relative to other surface regions. Starspots---and their bright counterparts, faculae---present a major challenge for exoplanet transmission spectroscopy \cite{Berdyugina_2005, Rackham_2019, Barclay_2021, Rackham_2022}. An exoplanet transiting a star with starspots present outside of the transit chord can create a wavelength-dependent distortion in the observed transit light curve, which can be misinterpreted as an atmospheric feature on the exoplanet. This can be particularly problematic for transmission spectroscopy, as the spectral features of starspots on cool stars can be very similar to those of some exoplanet atmospheres. For example, the presence of water vapor in the atmosphere of an exoplanet can create a similar absorption spectrum to a starspot that is also rich in water vapor \cite{Barclay_2021}. To combat this limitation, so that we may take advantage of the `small star opportunity' with {\it JWST}, we first need to better understand how stellar contamination affects transmission spectroscopy. 

\subsection{Pandora SmallSat} \label{sec:Pandora}

The goal of the Pandora mission \cite{Quintana_2021} is to disentangle stellar and planetary signals from transmission spectra in order to accurately determine the exoplanet's atmospheric composition. In support of this goal, Pandora has two major science objectives. The first is to measure the host star's spot and faculae covering fractions. To accomplish this objective, Pandora will observe transiting exoplanets using two wavelength channels simultaneously. One channel will capture the visible portion of the spectrum, covering from $\sim$0.4--0.8~$\mu$m through photometric observations. This channel will measure the time-varying stellar brightness which results from stellar inhomogeneity. The other channel will perform spectroscopy in the near-infrared (NIR), covering wavelengths from $\sim$0.9--1.7~$\mu$m. The spectroscopic observations from this channel will provide information about both the planet's atmospheric composition as well as variations in the stellar spectral signatures. Combining both the visible and NIR measurements improves our understanding of how the stellar host's surface inhomogeneities (from starspots and faculae) affect exoplanet transmission spectra.

The effects of stellar contamination on transmission spectra are a pressing concern in the exoplanet community\cite{Pont_2008, Pont_2013, Sing_2011, Rackham_2018, Rackham_2019, Rackham_2022}, particularly for targets around low-mass host stars, which tend to be more highly active. For this reason, Pandora's target list will focus on host stars that range from early-K to late-M stellar types. By limiting the range of host stars to these smaller, cooler stellar classes, Pandora will be able to provide population-level insights into stellar contamination effects on transmission spectra, not only from the worst-case offenders but also the classes of stars that are of high interest for characterizing small planets in the {\it JWST} era.

After correcting for stellar contamination, the second objective is to identify which exoplanets have hydrogen- or water-dominated atmospheres. One of the unexpected results from the \textit{Kepler} mission was the prevalence of sub-Neptune-sized planets. This discovery led to significant changes in our understanding of planet formation \cite[e.g]{Hansen_2012, Alibert_2013, Chatterjee_2014, Lee_2016}. Analysis of \textit{Kepler} data has also highlighted a transition point around 1.6 Earth radii, the so-called `Radius Valley' \cite{Fulton_2017}. Above this transition point, we see planets ranging in size from sub-Neptune to Jupiter that typically have lower mean-molecular-weight atmospheres that are both extended and hydrogen-dominated. Below this point are rocky planets that contain higher mean-molecular-weight atmospheres that are dominated by water or possibly heavier molecules. Several mechanisms currently work to explain this gap, including photoevaporation \cite{Owen_2013, Jin_2014} and core-powered mass-loss \cite{Gupta_2019}, though this remains a highly active area of research. Pandora's target list will include planets covering a wide range of sizes so that we can further understand how atmospheric composition varies with the planets' physical properties (e.g., radius and mass), the distance from its host star, and host star properties. By probing this parameter space, Pandora will be able to provide observational constraints that can help us to refine current theories on planetary atmosphere formation and evolution. 

Within this second objective for Pandora, there is an observational requirement to capture at least 10 transits for a minimum of 20 exoplanets during the year-long primary mission. This requirement places an upper bound on the orbital period of the planets on the target list to $\sim36$ days. The host star's brightness was also constrained in the selection of the notional targets to be between $6.5<~$J$~<11.5$.

With the parameter space established, a notional target list has been developed for mission planning purposes, which includes 20 host stars and 23 exoplanets. The notional list is provided in Table \ref{table:target_list}. During Pandora's science operations, the target list is intended to be highly flexible in order to respond to the rapidly growing field of exoplanet discovery.

\begin{table}
    \resizebox{\textwidth}{!}{%
    \begin{tabular}{|c|c|c|c|c|c|l|}
        \hline
        \textbf{Star Name} &
        \textbf{\begin{tabular}[c]{@{}c@{}}Spec. \\ Type\end{tabular}} &
        \textbf{\begin{tabular}[c]{@{}c@{}}J-band\\ Magnitude\end{tabular}} &
        \textbf{Planet} &
        \textbf{\begin{tabular}[c]{@{}c@{}}Planet Radius \\ (Earth Radii)\end{tabular}} &
        \textbf{\begin{tabular}[c]{@{}c@{}}Orbital Period \\ (days)\end{tabular}} &
        \multicolumn{1}{c|}{\textbf{\begin{tabular}[c]{@{}c@{}}Transit \\ Duration (hrs)\end{tabular}}} \\ \hline
        HAT-P-19   & K1   & 11.1 & b & 12.7 & 4.009 $\pm$ 6.00E-06  & 2.84 $\pm$ 0.0336   \\ \hline
        HAT-P-11   & K4   & 7.6  & b & 4.4  & 4.888 $\pm$ 1.30E-06  & 2.36 $\pm$ 0.0011  \\ \hline
        HIP 65 A   & K4   & 8.9  & b & 22.8 & 0.981 $\pm$ 3.10E-06  & 0.79 $\pm$ 0.006   \\ \hline
        GJ 9827    & K5   & 8    & c & 1.2  & 3.648 $\pm$ 6.33E-05  & 1.82 $\pm$ 0.0264  \\ \hline
        HATS-72    & K5   & 10.4 & b & 8.1  & 7.328 $\pm$ 1.60E-06  & 3.08 $\pm$ 0.0072  \\ \hline
        WASP-69    & K5   & 8    & b & 0.9  & 3.868 $\pm$ 2.00E-06  & 2.16 $\pm$ 0.0107  \\ \hline
        WASP-107   & K6   & 9.4  & b & 12.4 & 5.721 $\pm$ 2.00E-06  & 2.74 $\pm$ 0.0019  \\ \hline
        K2-141     & K7   & 9.1  & c & 7    & 7.749 $\pm$ 2.20E-04  & 2.21 $\pm$ 0.02    \\ \hline
        K2-3       & M0   & 9.4  & b & 2    & 10.055 $\pm$ 1.10E-05 & 2.56 $\pm$ 0.0390  \\ \hline
        GJ 3470    & M1.5 & 8.8  & b & 4.6  & 3.337 $\pm$ 3.90E-06  & 1.90 $\pm$ 0.0120  \\ \hline
        GJ 357     & M2.5 & 7.3  & b & 1.2  & 3.931 $\pm$ 8.00E-05  & 1.53 $\pm$ 0.12    \\ \hline
        GJ 436     & M2.5 & 6.9  & b & 4.2  & 2.644 $\pm$ 6.00E-07  & 1.01 $\pm$ 0.0034  \\ \hline
        LTT 1445 A & M2.5 & 7.3  & b & 1.4  & 5.359 $\pm$ 4.30E-06  & 1.37 $\pm$0.017    \\ \hline
        L 98-59    & M3   & 7.9  & c & 1.4  & 3.691 $\pm$ 2.60E-06  & 1.37 $\pm$ 0.017   \\ \hline
        L 98-59    & M3   & 7.9  & d & 1.6  & 7.451 $\pm$ 8.10E-06  & 0.527 $\pm$ 0.089  \\ \hline
        GJ 486     & M3.5 & 7.2  & b & 1.3  & 1.467 $\pm$ 3.10E-05  & 1.01 $\pm$ 0.1085  \\ \hline
        TOI 540    & M4   & 9.8  & b & 1.1  & 3.868 $\pm$ 2.00E-06  & 0.489 $\pm$ 0.1138 \\ \hline
        GJ 1132    & M4.5 & 9.2  & b & 1.1  & 1.629 $\pm$ 2.70E-05  & 0.75 $\pm$ 0.0964  \\ \hline
        LHS 1140   & M4.5 & 9.6  & c & 1.3  & 3.778 $\pm$ 3.00E-05  & 1.10 $\pm$ 0.025   \\ \hline
        LHS 3844   & M5   & 10   & b & 1.3  & 0.463 $\pm$ 1.90E-06  & 0.52 $\pm$ 0.0047  \\ \hline
        TRAPPIST-1 & M8   & 11.4 & b & 0.9  & 1.511 $\pm$ 6.00E-06  & 0.61 $\pm$ 0.0016  \\ \hline
        TRAPPIST-1 & M8   & 11.4 & e & 1    & 6.101 $\pm$ 3.50E-05  & 0.93 $\pm$ 0.0042  \\ \hline
        TRAPPIST-1 & M8   & 11.4 & f & 1.05 & 9.208 $\pm$ 3.20E-05  & 1.05 $\pm$ 0.0042  \\ \hline
    \end{tabular}}
    \vspace{.5em}
    \caption{Pandora's notional target list. The list covers targets spanning from Earth- to Jupiter-sized planets around a wide range of K and M stellar hosts. Due to observational requirements, the orbital period for the exoplanets was capped at 36 days.}
    \label{table:target_list}
\end{table}

\section{Methods} \label{sec:methods}

\subsection{Target Visibility} \label{subsec:target_vis}
To develop an observation schedule for Pandora, we start by assessing the visibility of each of the host stars within our notional target list, based on the Sun, Moon, and Earth avoidance angles. These avoidance angles are necessary to ensure light from the Sun does not damage the detectors while stray light from scattered and reflected light from the Earth or Moon does not cause significant degradation to the observation quality. For Pandora, the keep-out regions are a Sun-avoidance angle of $90^{\circ}$, a Moon-avoidance angle of $40^{\circ}$, and an Earth limb-avoidance angle of $20^{\circ}$. 

For target visibility, we also considered observations that would occur when Pandora crosses through the South Atlantic Anomaly (SAA). However, as discussed further in Sec. \ref{subsec:SAA}, we found that although the observational data collected during an SAA crossing is expected to have some degradation in quality, it will not prevent observations from being collected. For this reason, the SAA crossings are not considered in the visibility determination of the target and are only taken into account in the Quality Metric calculations.

To simulate the Pandora mission and orbit determination during its planned science observing year, we use the General Mission Analysis Tool (GMAT) \cite{Hughes_2014}, an open-source design and integration tool that models and optimizes spacecraft trajectories. 

\begin{figure}[h]
    \centering
    \includegraphics[width=\textwidth]{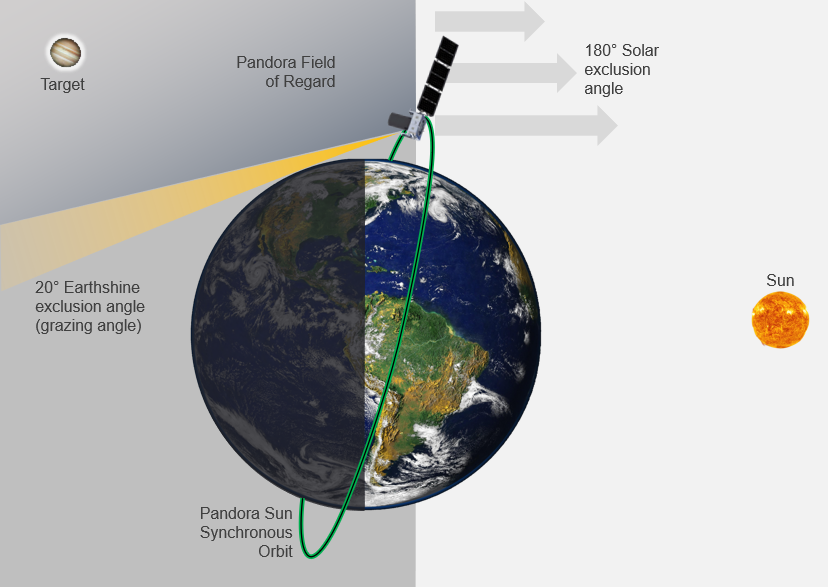}
    \caption{Illustration of Pandora's Sun-synchronous low Earth polar orbit, along with the depiction of the Earth and Sun exclusion zones. Not depicted is an additional exclusion angle with regard to the Moon of $40^{\circ}$. The Sun-synchronous orbit will allow Pandora to capture long-baseline observations of single targets while also providing a relatively stable thermal environment and excellent ground station access.}
    \label{fig:Orbit}
\end{figure}

As part of the NASA Astrophysics Pioneers Program under which Pandora was selected, NASA will provide the launch service through the Science Mission Directorate's rideshare program. This means Pandora will be a secondary payload, and an exact launch date has yet to be determined. Pandora will also not have an active propulsion system onboard, so its exact orbital parameters will not be known until after its orbital insertion. Therefore, for current planning purposes, we chose a launch date of March 25, 2025, based on when Pandora is required to be ready to launch and the likely rideshare options that currently exist around that time. With a planned commissioning period of one month, science observations will thus start on April 25, 2025. The orbital parameters that we have assumed for our GMAT simulation are shown in Table \ref{Orb_params} and are based on our current best estimates. Figure \ref{fig:Orbit} shows an illustration of Pandora's Sun-synchronous low Earth polar orbit with a local time of approximately 6~\textsc{AM}/6~\textsc{PM}. Using GMAT, we simulated Pandora's trajectory over the course of its planned 13-month in-orbit operations, with one month for commissioning and 12 months for science observations.

\begin{table}
    \centering
    \begin{tabular}{|l|c|c|c|c|c|}
    \hline
    \multicolumn{1}{|c|}{} & 
    \begin{tabular}[c]{@{}c@{}}Semi-Major \\ Axis\end{tabular} & Eccentricity & Inclination & \multicolumn{1}{c|}{\begin{tabular}[c]{@{}c@{}}Right Ascension of \\ the Ascending Node\end{tabular}} & \multicolumn{1}{c|}{\begin{tabular}[c]{@{}c@{}}Argument of \\ Periapsis\end{tabular}} \\ 
    \hline
    Pandora & 6921 km & 0.0 & $97.6^{\circ}$ & $238.7^{\circ}$ & $60^{\circ}$ 
    \\ \hline
    \end{tabular}
    \vspace{.5em}
    \caption{Pandora's estimated orbital parameters that were used in GMAT for orbital determination of Pandora during the planned science observing year. For the estimated launch date of March 25, 2025, 00:00:00 UTC, the True Anomaly for Pandora was set to $300^{\circ}$.}
    \label{Orb_params}
\end{table}

From GMAT we were able to extract two subsets of positional data. The first set was Earth-centric Cartesian coordinates for Pandora, the Sun, and the Moon. To calculate the positions of the celestial bodies, GMAT utilizes the NASA JPL Development Ephemerides, DE405 \cite{Standish_1998}. The second set provides Pandora's altitude, latitude, and longitude. Both sets were determined using a Dormand--Prince integrator \cite{Dormand_1980} within GMAT with a variable time step between 15 and 90 seconds. In order to simplify future calculations and standardize results, both data sets were then interpolated to a uniform time spacing of 60 seconds.

From the Cartesian coordinates data we calculated Pandora's pointing vector for each host star on our target list. With the pointing vector established, separation angles were then calculated with respect to each of the local bodies of interest (i.e., Sun, Moon, and Earth). By comparing the separation angles we calculated to the avoidance angles of the instruments, we created a target visibility calendar for all of our host stars.

\begin{figure}
    \centering
    \includegraphics[width=\textwidth]{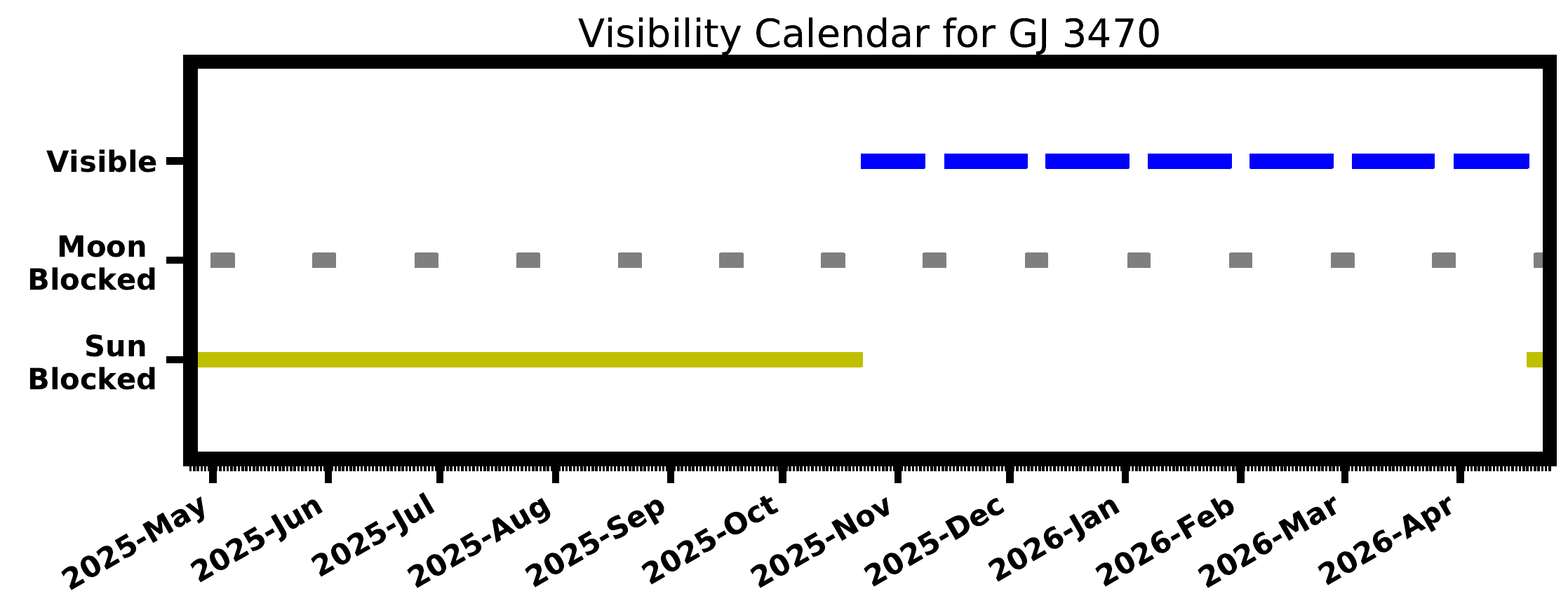}
    \caption{Target visibility calendar for host star GJ~3470 for an observation year from April 25, 2025, to April 25, 2026. The yellow bar depicts the time during the year when the target is within the exclusion angle of the Sun and is therefore not visible to Pandora. The grey bar depicts similar visibility constraints with respect to the Moon. Here we can see the 28-day cycle of the Moon moving in and out of the field of view for our target. Also illustrated is the Sun-blocking geometry that covers half the year due to the angular separation restriction imposed of $90^{\circ}$ between the target, Pandora, and the Sun. The total visibility (blue bar), accounts for the blocking from all sources of interest. Due to the timing resolution of the graph, the times where GJ~3470 is blocked from Pandora's view by the Earth are not visible. Instead, Fig. \ref{fig:Vis_cal_earth} illustrates the cadence at which Earth's blocking becomes relevant.}
    \label{fig:Vis_cal}
\end{figure}

\begin{figure}[h]
    \centering
    \includegraphics[width=\textwidth]{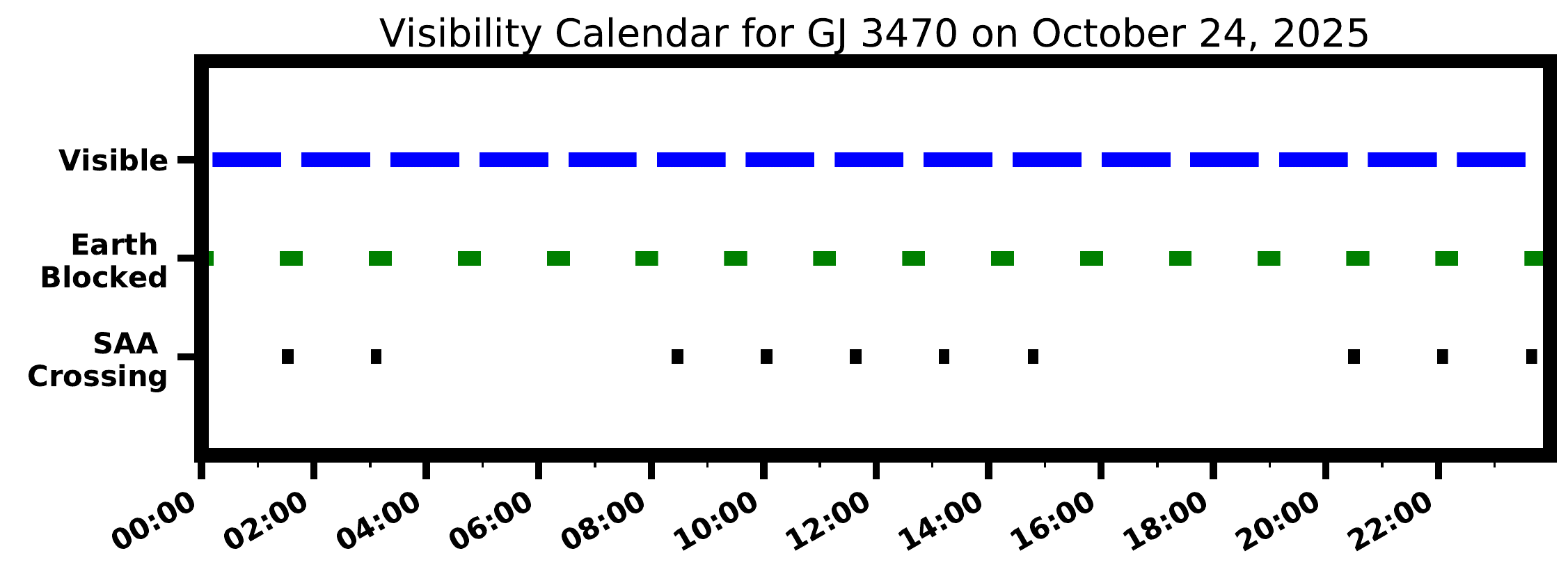}
    \caption{Target visibility for host star GJ~3470 for October 24, 2025. At this temporal resolution the gaps in visibility due to the target being inside the Earth limb-avoidance angle of $20^{\circ}$ (green bar) are visible. On this day, the target is entirely visible with respect to the Moon and Sun, so the total visibility of the target (blue line) is strictly dictated by Earth--Pandora--GJ~3470 geometry. On average over the observing year, the Earth blocks about 20 minutes per 95.5-minute orbit. At this temporal resolution, the frequency of SAA crossings by the observatory is also visible. Although the observational data collected during an SAA crossing is expected to have some degradation in quality, it will not prevent observations from being collected. For this reason, the SAA crossings are not considered in the visibility determination of the target and are only taken into account in the Quality Metric calculations.}
    \label{fig:Vis_cal_earth}
\end{figure}

Figure \ref{fig:Vis_cal} is an example visibility calendar for GJ~3470, showing gaps in the observing window due to Moon- and Sun-avoidance constraints. At this resolution, the gaps caused by Earth-limb avoidance are obscured. With an estimated altitude of 550~km, Pandora will have an orbit of about 95.5 minutes. Based on analysis of the notional target list, under the worst viewing conditions identified, the Earth exclusion angle will block roughly 35 minutes of a single orbit, leaving 60 minutes of target visibility. On average, though, most targets will be visible for 65--75 minutes per orbit. In Fig. \ref{fig:Vis_cal_earth} we highlight a single day during the observing calendar where neither the Sun nor the Moon block observations of GJ~3470. Figure \ref{fig:Vis_cal_earth} illustrates the higher cadence observation gaps that are due to Earth-limb avoidance geometry arising from Pandora's Sun-synchronous polar orbit.

\subsection{Quality Metric} \label{subsec:qfactor}
To optimize the science yield of Pandora, we developed a metric to quantify the expected science quality of an individual transit event. Although the quality metric presented herein has been made with the Pandora Smallsat mission in mind, it is also general enough for use with other Earth-orbiting observatories interested in transiting exoplanets. For our quality metric, we consider the observing efficiency of the observatory as a weighted factor, which we describe in Sec.~\ref{subsec:efficiency}. Although this factor does not directly influence the quality of our individual transit observations, it does influence the total science yield of the mission. Beyond meeting the required science goals of the mission, creating an efficient schedule will allow Pandora to conduct more observations, whether it's revisiting particularly interesting targets, observing additional new targets, or conducting auxiliary science cases. We have also added two additional considerations that directly affect the scientific return of each individual transit. Both considerations are discussed in greater detail below in Secs.~\ref{subsec:transit_cover}, and \ref{subsec:SAA}. In brief, these two additional factors are 1) how much of the transit can be observed by Pandora and 2) how much (if any) of a transit occurs during an SAA crossing by Pandora.

\subsubsection{Observing Efficiency} \label{subsec:efficiency}
\begin{figure}[h]
    \centering
    \includegraphics[width=\textwidth]{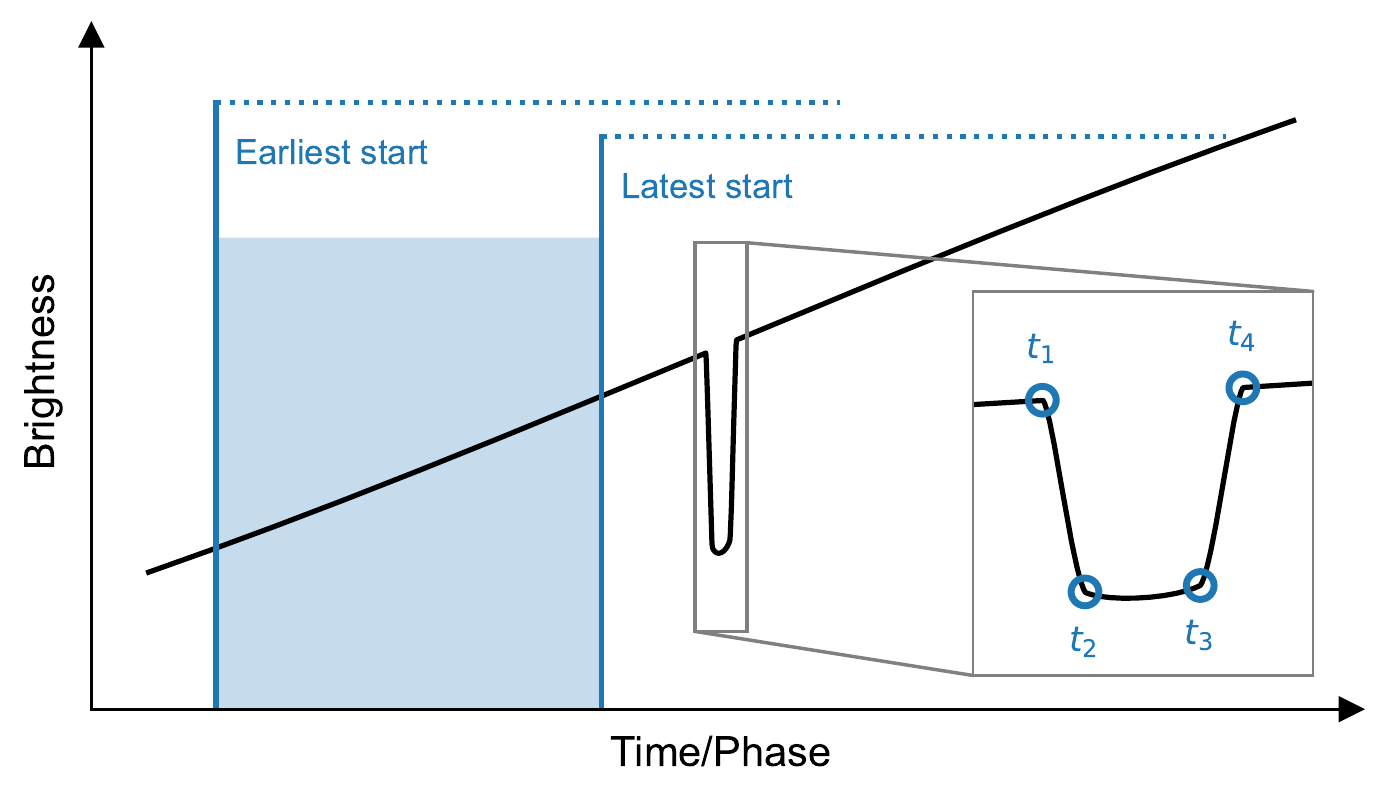}
    \caption{Diagram of a transit light curve (solid black line), along with the allowable range for the observation window within the scheduler. The transit light curve was created with batman \cite{Kreidberg_2015}, using a host star input that includes a sinusoidal stellar variability with a period of 10 days. The inset enlarges the exoplanet transit, highlighting key transit contact points: t$_1$ and t$_2$ show the start and stop of ingress as the planet moves in front of the host star, while t$_3$ and t$_4$ show the start and stop of egress as the planet moves out of the stellar disk. The light blue shaded region before the transit is the observing start window we implement for scheduling. The `earliest start' ensures we have a minimum of 4 hours of out-of-transit observation time following the end of egress (t$_4$) before the end of our fixed 24-hour observation of a target. Conversely, the `latest start' ensures we have a minimum of 4 hours of out-of-transit observation time prior to the start of ingress (t$_1$). The horizontal dashed  lines show the extent of the 24 hours of observations for the earliest and latest start conditions.}
    \label{fig:transit_diagram}
\end{figure}

Pandora's planned observing strategy will be to select a target and observe it for a 24-hour period. During that 24 hours, we will capture the transit of the planet of interest. The transit duration for targets on our notional list ranges from 0.5 to 3 hours, with an average duration time of 1.5 hours. The remaining observing time is used to gather long baseline measurements of the host star outside of the transit. The out-of-transit measurements will be used to better understand the photospheric heterogeneity of the host star. Once characterized, the effects of stellar photospheric heterogeneity on the extracted planet spectrum can be properly corrected. We also want to ensure we have out-of-transit measurements on both sides of the exoplanet transit. To accomplish this we defined an observing start window for each transit with an early and late start, schematically depicted in Fig. \ref{fig:transit_diagram}. The early start is defined as the earliest time we could begin observing in a 24-hour window to guarantee 4 hours of out-of-transit coverage after the transit ends (t$_4$). The late start is conversely defined as the latest time we could begin observing to guarantee 4 hours of out-of-transit coverage before the transit begins (t$_1$).

With this observing strategy in mind, we determine the observing efficiency of each transit by calculating the amount of time between the end of the previously scheduled event and the `early start' for the transit. This gap is then divided by our baseline 24-hour observing window and subtracted from 1. We restrict our observing efficiency calculations to only those transits that could start within the next 24-hour window following the last scheduled event. This allows our observing efficiency factor to naturally fall between 0 (low efficiency) and 1 (high efficiency).

\subsubsection{Transit Coverage} \label{subsec:transit_cover}

The next contribution to the quality factor considers how much of a planet's transit is able to be captured by Pandora. For this factor, we start by calculating the start and end time (t$_1$ and t$_4$ contacts in Fig. \ref{fig:transit_diagram}) for each transit that will occur during the observing year. We then use those times to construct a transit window with a 1-minute temporal resolution for each transit. Next, we cross-check each of these potential transits against the target's visibility calendar to see if there is an overlap between the transit and any keep-out region restrictions. If any overlap exists, then that overlap period is considered `blocked' since Pandora would be unable to observe the target during that period. 

We also consider visibility-blocking events that can occur due to other known planets in the target's host system. As an extreme example, the TRAPPIST-1 system has seven known planets; therefore, in order to calculate the transit coverage for TRAPPIST-1b, we also calculate all the possible transits of TRAPPIST-1c through TRAPPIST-1h and look for overlaps with TRAPPIST-1b transits. 

\begin{figure}[h]
    \centering
    \includegraphics[width=\textwidth]{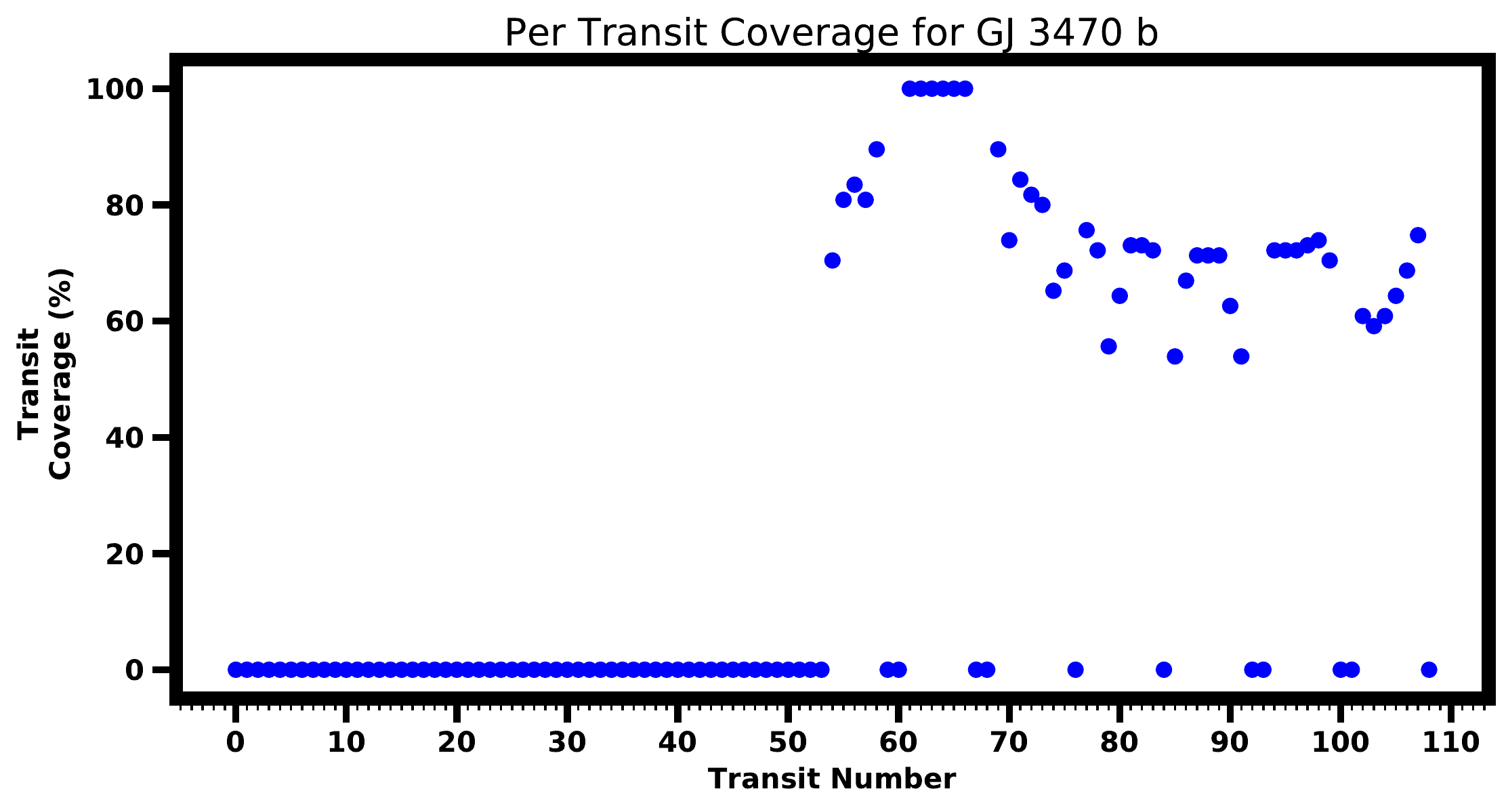}
    \caption{Over the course of Pandora's planned observing year, GJ~3470b will transit its host star 109 times. This graph shows each transit during that year and the percentage of each event Pandora is capable of capturing after considering target visibility. When compared to the visibility calendar in Fig. \ref{fig:Vis_cal} and \ref{fig:Vis_cal_earth}, we see that in roughly the first half of the year, there is no transit coverage due to the keep-out angle with the Sun. There are also seven total-blocking periods (i.e. 0\% transit coverage) in the second half of the year that are caused by the keep-out angle of the Moon. The remaining visible transits vary in coverage largely due to Earth occultations from Pandora's orbit during the transit and a few partial Moon occultation events.}
    \label{fig:Transit_coverage}
\end{figure}

Once all observation blocking conditions are assessed we calculate the fraction of the transit window that is able to be captured by Pandora. This factor ranges from 0, meaning none of the transit can be observed, to 1, meaning the entire transit can be observed. Figure \ref{fig:Transit_coverage}, shows the transit coverage for each transit of target GJ~3470b over the science observing year.

\subsubsection{SAA Crossing Coverage} \label{subsec:SAA}

For the final quality factor contribution we consider the effect of the SAA on our science data. The SAA is the near-Earth region where Earth's magnetic field is particularly weak and therefore Earth's inner Van Allen radiation belt reaches the closest to Earth's surface ($\mathord{\sim}200$~km). Because of the weakened magnetic field in this region, spacecraft whose orbits cross the SAA experience increased levels of high-energy radiation while traversing this zone. The increased radiation can cause a wide range of effects on scientific instruments, from short-term effects like single-event-upsets in the electronics or high cosmic ray counts on a detector, to long-term effects like pixel charge persistence.

To assess the likelihood and severity of these different events on the Pandora detectors, we looked at results from some prior and existing missions with similar detectors. For Pandora's near-infrared observations, the satellite will be using a HAWAII-2RG (H2RG) detector, a flight spare from {\it JWST}'s NIRCam instrument. Unfortunately, we can not assess SAA effects on the H2RG using data from {\it JWST} since it is not in LEO and therefore does not have to worry about the effects of the SAA. However, we can use data from {\it HST}, which is in LEO and whose Wide Field Camera 3 (WFC3) utilizes an earlier version of the detector, the H1RG. Earlier work has shown that although observations can be taken during an SAA crossing with the WFC3/IR H1RG detector, it results in $\sim$12\% of pixels being affected by cosmic rays per minute \cite{Barker_2009}. The persistence of cosmic rays on the detector decay following a power law; while the most energetic cosmic rays last longer, up to an hour, the majority of cosmic rays only affect the detector on the order of tens of minutes. Not only are the effects short-lived but in several studies where exoplanet transits are observed during SAA crossings, we see that in most cases the cosmic rays can be identified and corrected with some effort \cite[e.g]{de_Wit_2018, Zhang_2018}. 

For its visible-wavelength photometry and a component of its pointing control, Pandora will use a PCO Panda 4.2M sCMOS sensor. This is similar in design and bandpass to the CMOS imager that was used on the ASTERIA CubeSat. ASTERIA was also in LEO and took observations even while it crossed through the SAA. One such crossing occurred during an observation of a secondary eclipse of 55 Cancri e. In their analysis of that data, Knapp et al. \cite{Knapp_2020}, found that, as expected, the detector did experience an increase in hot pixels caused by cosmic rays. These transient hot pixels, however, were found to only last a short time (one or two 50~ms frames) and caused negligible disruption to their pointing stability. Moreover, when compared to their 1-minute coadded exposures, it was determined that these transient events did not end up affecting their photometric precision \cite{Knapp_2020}. 

Based on these findings, we determined that an SAA crossing will not have significant effects with regard to our out-of-transit observations, since we plan to acquire at least four hours of observations on either side of a transit event. Therefore, we only considered SAA contamination to be detrimental to the scientific return for observations in which the SAA crossing occurs within a planet's transit, that is, between t$_1$ and t$_4$ in Fig. \ref{fig:transit_diagram}. Following the SAA contours in Ref. \citenum{Barker_2009} for WFC3/IR, we define the boundaries of the SAA for Pandora to be a box that covers latitudes $-40^\circ$ to $0^\circ$ and longitudes $-90^\circ$ to $30^\circ$.

To quantify this degradation in science return due to the SAA, we determined when Pandora would be within the SAA boundary area using Pandora's latitude and longitude positions as determined by GMAT. We then calculated the fraction of overlap between each transit, using the transit window developed for the transit coverage calculation and any SAA crossing that occurs during that window. This calculation results in a value between 0 (no overlap) and 1 (fully overlapping). 

\subsection{Minimum Transits Requirement Metric} \label{subsec:MTRM}
As mentioned previously, one of the observational requirements of the Pandora mission is to observe a minimum of 20 planets, and for each planet observe at least 10 transits. To address this mission requirement we instituted an additional component to our scheduler that can override our quality metric. We call this component the Minimum Transits Requirement Metric or MTRM. To calculate the MTRM for each planet, we determine how many transits are left in the observing year for that planet. We then utilize the transit coverage factor calculated previously to remove any transits from the list that drop below a minimum threshold (e.g., 25\% transit coverage). This provides the number of possible transits left during which Pandora can observe the target. We then divide the possible transits left by the number of transits still necessary to reach the 10-transit minimum, and this becomes our MTRM. If during the scheduling sequence, it is determined that one or more of the targets schedulable in the next 24-hour window have an MTRM~$<~2$, then the target with the lowest MTRM is immediately scheduled regardless of how it ranked with regard to the quality factor score.

\section{Results} \label{sec:results}
\begin{figure}[htbp]
    \centering
    \includegraphics[width=\textwidth]{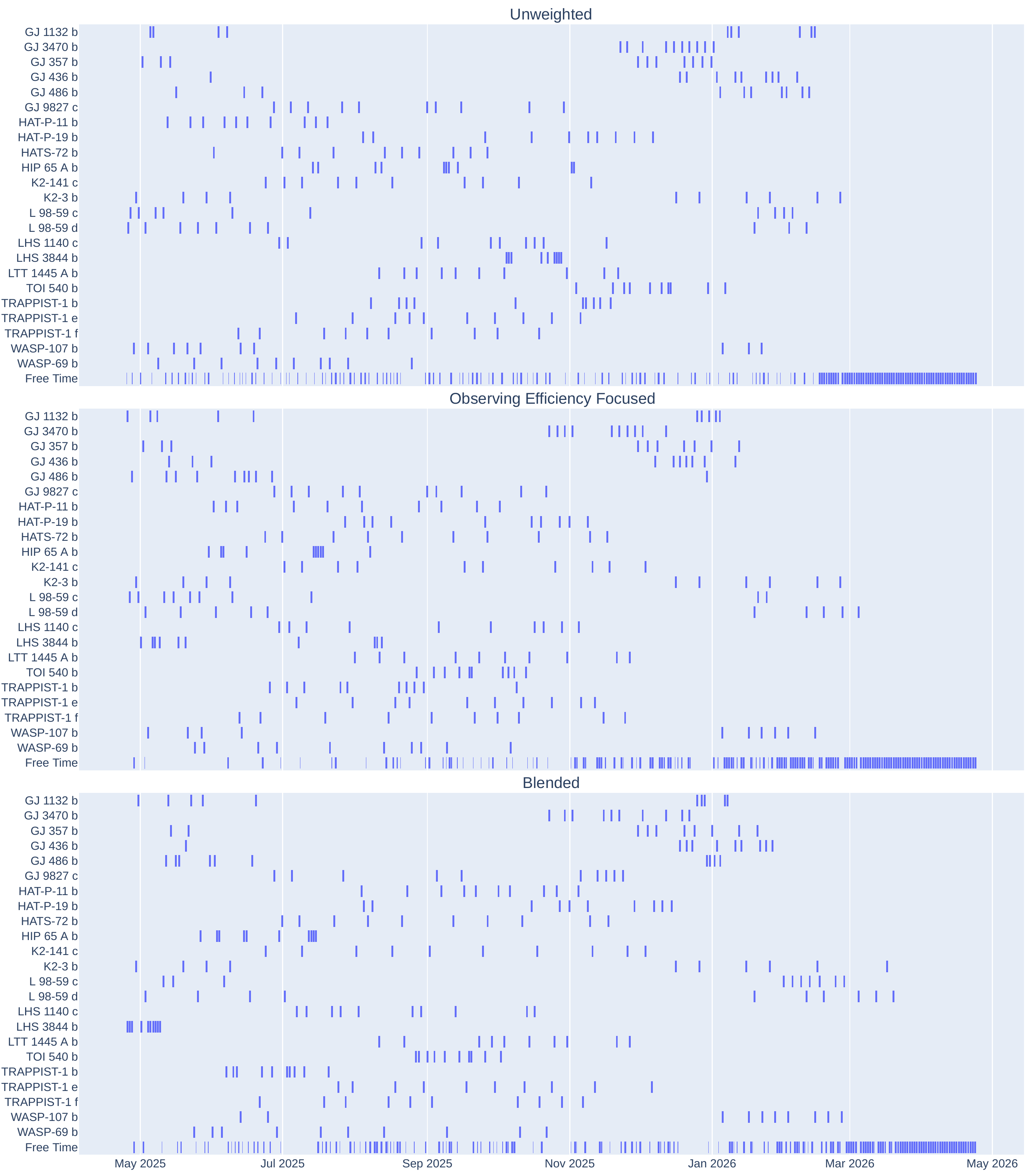}
    \caption{The three schedules built using Pandora's notional target list, which serves as our comparison. The first schedule, labeled `Unweighted', has no weights given to any of the quality factors. The second schedule, labeled `Observing Efficiency Focused', is designed to strictly optimize the observing efficiency of Pandora. The third schedule, labeled `Blended', has a weighting strategy that places 25\% into observing efficiency, 50\% toward optimizing transit coverage, and 25\% to minimizing SAA overlaps. The observation year for all three schedules goes from a start date of April 25, 2025, to April 25, 2026.}
    \label{fig:Opt_schedule}
\end{figure}

To investigate the effects the schedule can make on the factors of interest we compare three different schedules. The first is an unweighted schedule made prior to implementing any of the quality factors discussed above; this program simply scheduled whichever target had the least number of visible transits left in Pandora's lifetime at the time of each scheduling window. This provided a quick test to confirm that we would be able to, at a minimum, observe all the targets on our list for the required 10 transits each. We next developed a schedule that simply aimed to minimize the wait times between our 24-hour observation sets and included the MTRM to ensure the capture of 10 transits per target. This scheduling scheme was blind to the two new science quality factors: Transit Coverage and SAA Overlap. The final schedule considers all three quality factors, with a greater focus on transit coverage. In this program, we distributed the weights for the three quality factors so that a quarter of the effort went to optimizing the observing efficiency, half of the effort went to maximizing the transit coverage in each observation, and the remaining quarter went to minimizing SAA crossing overlap with the transits. The resulting schedule built from this final method is shown in Fig.~\ref{fig:Opt_schedule}. 

\begin{figure}[h]
    \centering
    \includegraphics[width=\textwidth]{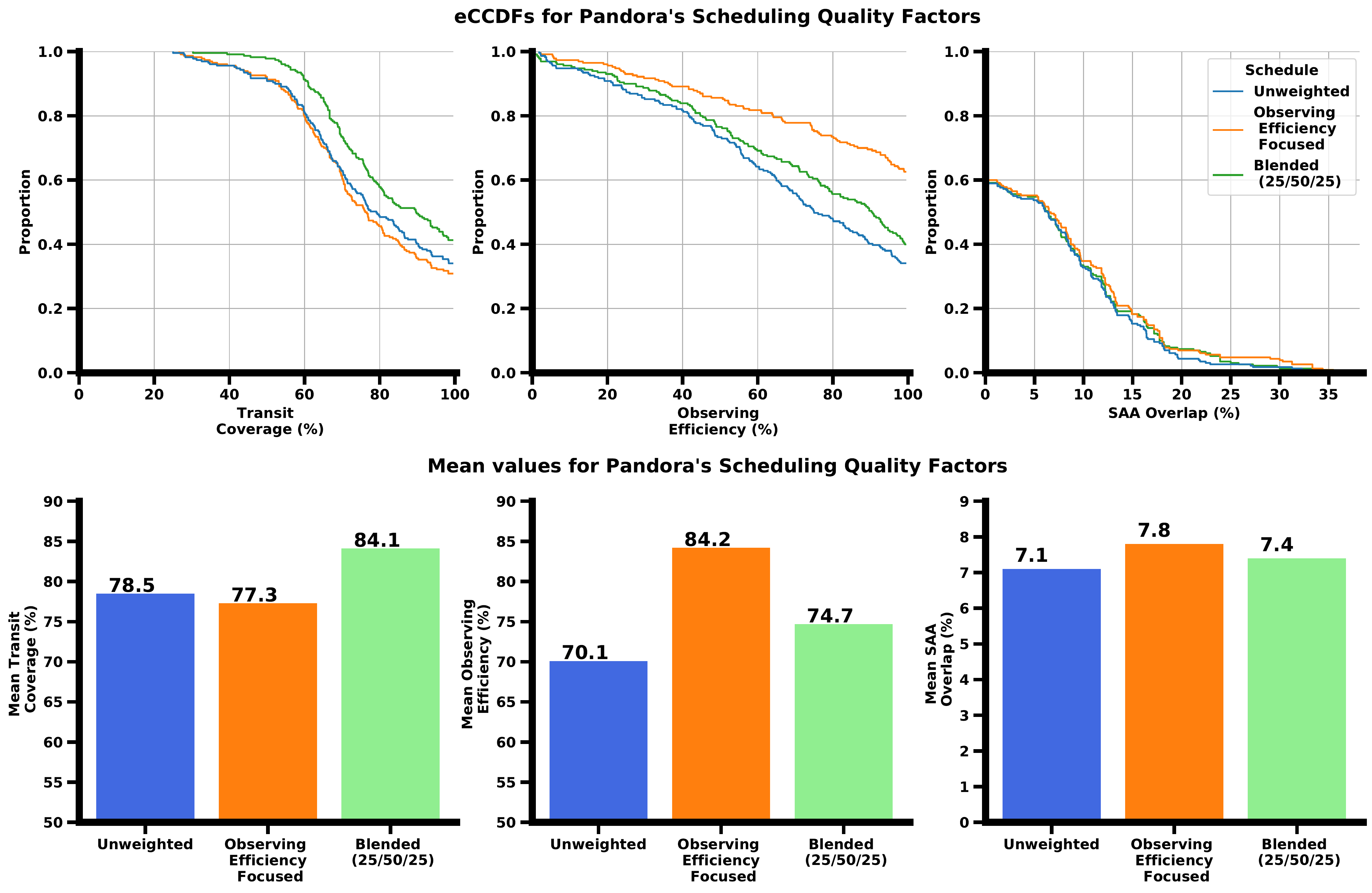}
    \caption{Comparison of the three quality factors for the three different schedules. The top row compares each schedule across the three quality metrics using empirical complementary cumulative distribution functions (eCCDF). The eCCDFs show the total proportion of all 230 transits captured by each schedule that are at or above a given quality factor percentage.
    The bottom row illustrates the mean values of the quality factors for each schedule, highlighting the trade-off between optimizing observing efficiency and optimizing transit coverage. Also highlighted is the negligible difference seen in the SAA overlap regardless of the weighting strategy.}
    \label{fig:quality_ecdf}
\end{figure}

With all three schedules, we were able to successfully capture the 230 necessary transits to meet the observational requirement of 10 transits per target using our notional target list. Where we begin to see differences in the three schedules is when we delve into the quality factors, highlighted in Fig.~\ref{fig:quality_ecdf}. In the top row of Fig.~\ref{fig:quality_ecdf} are empirical complementary cumulative distribution functions (eCCDF), which show the proportion of the 230 transits captured in a given schedule that meets the quality metric value on the x-axis. The bottom row of Fig.~\ref{fig:quality_ecdf} shows the mean value for each quality factor across all 230 transits of each schedule.

The factor with the smallest difference across the three schedules is the SAA overlap. In Fig.~\ref{fig:quality_ecdf}, we see that, regardless of the weighting strategy chosen, the SAA Overlap follows the same basic cumulative distribution. We find that across the three schedules, the average SAA overlap per transit is between 7--8\%. This is in line with the total proportion of time Pandora will be in the SAA region during its observing year, which based on our estimated orbital parameters (Table \ref{Orb_params}), is found to be 7.5\% of the year. The lack of variance in this metric can be attributed to the fact that the SAA crossings are a function of the orbital geometry of Pandora and are therefore fixed in time regardless of the target. These crossings, although short in duration ($\sim$10~minutes), also occur at high enough frequency, illustrated in Fig.~\ref{fig:Vis_cal_earth}, as to be unavoidable. The only way the scheduler could significantly improve this metric, and reduce SAA overlap with transit observations, is by not observing during SAA crossings; however, doing so would jeopardize the mission's requirement of capturing 10 transits per target. 

A greater divergence between the blended schedule and the other two schedules emerges when considering transit coverage. For this quality factor, we see in Fig. \ref{fig:quality_ecdf} that both the observing-efficiency-focused and the unweighted schedules are very similar in their distribution, as would be expected with neither placing an emphasis on this factor. There is a slight separation in transit coverage in the $\sim$~80--100\% coverage region, with the unweighted having a slightly higher number of transits covered at a full 100\%; but on average this difference is negligible, with both having an average per-transit coverage between 77--78\%. The blended schedule, on the other hand, placed high importance on this quality factor and has a lower proportion of its transits in all coverage percentages except for full transit coverage. In this category, we can see this in the fact that the entire eCCDF curve for the blended schedule is shifted to the right compared to the other schedules. Not only does the blended schedule improve the total number of transits captured with 100\% coverage by about 10\%, but its average per transit coverage is also improved by $\sim$5\%. 

The improved transit coverage does come at a trade-off, however, with the observing efficiency of the schedule. This can be seen clearly in Fig.~\ref{fig:quality_ecdf}, especially in comparing the blended and the observing-efficiency-focused schedules. As expected, the schedule that was focused on maximizing the observing efficiency does the best in this quality factor. On a per-transit average, we find that the observing efficiency achieved by this schedule is over 10\% higher than the blended schedule and more than 15\% better than the unweighted schedule.

\section{Conclusion} \label{sec:conclusions}
In this work, we explore ways to optimize exoplanet transit spectroscopy observations beyond the traditional method of optimizing observing efficiency. To this end, we considered two additional factors that directly affect the quality of our science data, the total amount of a transit captured during an observation and how much of the transit data is contaminated due to the observation being recorded during an SAA crossing by the satellite. We present the framework used to develop a scheduling algorithm that allows for the adjusting of the importance of the three quality factors: observing efficiency, transit coverage, and SAA overlap. We conclude by presenting a comparison of a few different scenarios that may be considered when building the `optimal' schedule.

We confirm that regardless of the scheduling strategy used to develop the baseline year schedule for the Pandora SmallSat mission, the final schedule will be able to successfully meet the observational requirement of capturing at least 10 transits for each of the targets based on the notional target list. This success is in part due to the MTRM feature which acts as a safeguard so that preferences do not interfere with achieving mission success. This ensures that the quality factor weightings can be adjusted to reflect what the team determines to be important while instilling confidence that the observational requirements are met. In our comparison of the different schedules built, we found that the SAA overlap quality factor was the least affected by changes to the weighting efforts, showing negligible improvement when taken into consideration. We also explored the trade-off between optimizing observing efficiency and transit coverage.

\section{Future Development} \label{sec:future}
One current feature that needs to be enhanced is the ancillary science event scheduling component. This feature currently allows users to list desired science cases they wish to add to the schedule. These events may not be phase-constrained but instead constrained in time like a transient event. In the current form, these ancillary science events are listed in a CSV file along with any necessary time constraints. The scheduler then uses those time-constraint windows to place the events in the schedule. The issue is that this feature can supersede the MTRM during the scheduling process so there is a chance that it will overwrite a transit that would otherwise need to be scheduled in order to meet the 10-transit requirement for a given planet. 

This problem will be addressed with two additional steps in the scheduling process. The first will be for the scheduler to attempt to block out the timing windows for all ancillary events and determine if it is still possible to schedule all the necessary planetary transits. If it is unable to accomplish this step successfully, then the second step will be implemented, which will force the schedule to have the transit take precedence and raise a flag to inform the user of a conflict between scheduling the ancillary event and accomplishing the main science goal.

An additional feature that will need to be added to the scheduling algorithm is a more thorough integration of operations activities into the schedule, including the scheduling of ground-station contacts. With Pandora planned to be in a polar orbit, it was chosen to use two ground stations operated by Kongsberg Satellite Services (KSAT): the Svalbard Satellite Station in Svalbard, Norway, and the Troll Satellite Station in Queen Maud Land, Antarctica. Knowing our ground station locations allows us to tap into another functionality of the GMAT software, which has the ability to calculate ground station contact opportunities. By providing the location and minimum elevation angle for each ground station of interest, GMAT will calculate the times and durations of possible contacts between the spacecraft and the ground station during the spacecraft's lifetime. By utilizing these output files we will be able to inform the Pandora scheduler when potential ground station contacts are available to schedule. 

The next step will be to determine when Pandora needs to make contact with the ground stations. From preliminary assessments from our data management, we expect to need three ground station downlinks each day, each lasting 8--10 minutes in duration. During the data management analysis, the Pandora team developed a couple of different observing schemes, with differing data volume production, that we can change depending on the brightness of the target. Knowing our data production rate for each target, we will be able to have the scheduler keep both a running data budget as well as an anticipated budget. It can then compare the data budget with our data limits and with the ground station contact opportunities to optimally schedule ground station contacts during each 24-hour window of observations while minimizing impacts on our science quality.

Continuing along the lines of incorporating operations activities, another improvement that will be made to the scheduler will be a more robust calculation of the observing strategy that takes into account the slewing time needed to move from one target to the next. One possible outcome from this more detailed accounting is an observing schedule that more efficiently schedules groups of targets that are close together in the sky. This grouping strategy would minimize the overhead time associated with slewing between targets and also reduce the overall demand on Pandora for momentum management. By minimizing these two operations activities, we would be able to increase the amount of time spent on targets, thus providing the scheduler with another way to increase observing efficiency.


\subsection*{Disclosures}
The authors have no relevant financial interests in the manuscript and no other potential conflicts of interest to disclose.

\subsection*{Acknowledgments}
T.F. is supported by the National Aeronautics and Space Administration (NASA) FINESST grant 80NSSC22K1893. Pandora is supported by NASA’s Astrophysics Pioneers Program.

\subsection* {Code, Data, and Materials Availability} 
The scheduler code is open-source and available on GitHub at: \newline \hyperlink{https://github.com/PandoraMission/pandora-scheduler}{ https://github.com/PandoraMission/pandora-scheduler}


\bibliography{report}   
\bibliographystyle{spiejour}   


\vspace{2ex}\noindent\textbf{Trevor Foote} is currently a Ph.D. candidate in the Astronomy Department at Cornell University. He received his bachelor's degrees in civil engineering and astrophysics from Washington State University and his master’s degree in astronomy from Cornell University. His research interests include mission planning and detector development, specifically for exoplanet atmospheric characterization science.

\vspace{1ex}
\noindent Biographies and photographs of the other authors are not available.

\listoffigures
\listoftables

\end{spacing}
\end{document}